\documentclass{aa}
\usepackage{graphicx}
\usepackage{txfonts}
\usepackage{natbib}
\bibpunct{(}{)}{;}{a}{}{,}
\begin{document}

\title{Partially-erupting prominences: a comparison between observations and model-predicted observables}

\author{D. Tripathi\inst{1}, S. E. Gibson\inst{2}, J. Qiu\inst{3}, L. Fletcher\inst{4}, R.Liu\inst{5}, H. Gilbert\inst{5}, H. E. Mason\inst{1}}

\offprints{D.Tripathi@damtp.cam.ac.uk}

\institute{Department of Applied Mathematics and Theoretical Physics, Wilberforce Road, Cambridge, CB3 0WA, UK\\ 
    \email{[D.Tripathi; H.E.Mason]@damtp.cam.ac.uk} 
    \and High Altitude Observatory, National Center of Atmospheric Research, Boulder, Colorado, USA\\
    \email{sgibson@ucar.edu} 
    \and Montana State University, Bozeman, MT 59717-3840, USA\\ 
    \email{qiuj@mithra.physics.montana.edu} 
    \and University of Glasgow, Glagow G12 8QQ, UK\\
    \email{lyndsay@astro.gla.ac.uk} 
    \and Department of Physics and Astronomy, Rice University, 6100 Main St., Houston, TX 77005, USA\\ 
    \email{[hgilbert; rliu]@rice.edu} }

\date{Received(date); Accepted(date)}

\abstract{}{To investigate several partially-erupting prominences to study their relationship with 
other CME-associated phenomena and to compare these observations with observables predicted by a model 
of partially-expelled flux ropes (Gibson \& Fan, 2006a, b).}{We have studied 6 selected events with 
partially-erupting prominences using multi-wavelength observations recorded by the Extreme-ultraviolet 
Imaging Telescope (EIT), Transition Region and Coronal Explorer (TRACE), Mauna Loa Solar Observatory 
(MLSO), Big Bear Solar Observatory (BBSO) and soft X-ray telescope (SXT). The observational features 
associated with partially-erupting prominences were then compared with the predicted observables from 
the model.}{The partially-expelled-flux-rope (PEFR) model of Gibson \& Fan (2006a, b) can explain the 
partial eruption of these prominences, and in addition predicts a variety of other CME-related observables 
that provide evidence for internal reconnection during eruption. We find that all of the 
partially-erupting prominences studied in this paper exhibit indirect evidence for internal reconnection. 
Moreover, all cases showed evidence of at least one observable unique to the PEFR model, e.g., dimmings 
external to the source region, and/or a soft X-ray cusp overlying a reformed sigmoid.}{The PEFR model 
provides a plausible mechanism to explain the observed evolution of partially-erupting-prominence-associated 
CMEs in our study.}


\titlerunning{Partially erupting prominences}
\authorrunning{Durgesh Tripathi et al. }

\maketitle

\section {Introduction}

Coronal mass ejections (CMEs) are routinely interpreted as possessing a helical magnetic flux rope 
structure \citep[see e.g.][]{chen_97, dere, plunkett}. Magnetic clouds, i.e., interplanetary structures 
that have been shown to be associated with CMEs, are also interpreted to be magnetic flux 
ropes \citep[][]{burlaga1981, burlaga82, Burlaga88}. An ongoing controversy remains, however, as to 
whether a precursor flux rope exists as a coronal equilibrium state prior to eruption, or whether it is 
formed during eruption. This is an important question to resolve, since CME initiation models and space 
weather predictions depend upon a clear understanding of the configuration of pre-CME magnetic fields and 
their evolution during eruption.

The existence of a precursor magnetic flux rope is an attractive concept from a theoretical point of view, 
as it may represent a minimum magnetic energy configuration \citep{taylor,low96,low99,rust2003,janse}. A 
precursor flux rope has also been used to explain a wide range of pre-CME phenomena, including photospheric 
magnetic flux evolution \citep{litesetal,lopezfuentes,green_00,fan01,mandrini_01,gibetal_04} and 
prominences, associated white-light cavities, and soft X-ray 
sigmoids \citep{priest,rk94,auldem_98,amari_99,gibetal_04,vanball_04, gibfan_06b}. For CMEs, where white-light, 
low coronal observations are available and which are unobscured by unrelated features along the line of sight, 
the prominence and its cavity have been tracked from pre-eruption through their expansion outwards in the CME
\citep{fisher,illhund85,hundface,sriva,maricic,gibcav}. In many cases, a flux rope model can be used to match 
the magnetic flux and chirality of precursor structures to magnetic 
clouds \citep{bothmerrust_97,bothmerschwenn_97,rust_05}.

On the other hand, observations associated with flares demonstrate the importance of magnetic reconnection in 
the eruption, and challenge the picture of a pre-existing flux rope simply losing equilibrium and expanding out 
into interplanetary space. Such a purely ideal eruption is inconsistent with observations which find that 
the impulsive stage of the flare is linked to that of maximum CME acceleration, implying that reconnections 
are significant to the dynamic evolution of the CME \citep{zhang01,trip_06a,chifor_06,zhangdere,chifor_07}. 
Most models of an erupting (pre-existing) flux rope do involve significant magnetic reconnections at a current 
sheet below the rope which serve to 'close down' the field beneath it \citep[e.g.,][]{linetal}, and are 
consistent with post-eruption arcade formations and flare-CME timing observations 
\citep[see e.g.,][]{trip_04, tripathi}. However, the field lines of the rope itself may not be involved in these
reconnections (see \cite{amari_03b} for an exception). Observations of soft X-ray loops, as well as of 
chromospheric flare ribbons, indicate that reconnection occurs initially along highly sheared loops, and
only later do the magnetic field lines make a transition to more potential arcade 
loops \citep{canfield,martinmcal,suetal_2006a,suetal_2006b} in a manner explained by models where reconnections 
take place initially in a sheared magnetic core \citep[e.g.,][]{moore_97}. A further analysis of flare ribbons 
implies that the bulk of magnetic cloud poloidal flux originates in reconnecting field lines \citep{qiuetal_06}, 
and studies of magnetic cloud charge states indicate possibly flare-associated heating along 
prominence-mass-carrying field lines \citep{skoug,gloeck,reinard}. These observations indicate that the
flux rope that escapes in the CME is made up of field lines that have undergone significant reconnection, as would 
be the case if the flux ropes were formed {\it in situ} during eruption, but which would not be the case for 
a pre-existing rope expanding in its entirety upwards without significant change of topology/connectivity.

Neither a totally erupting, pre-existing rope, nor a rope that forms completely {\it in situ} during eruption 
can explain the full range of these observations. An alternative to either of these extremes is a
'partially-expelled flux rope' (PEFR) model \citep{gibfan_06a}. In this model, a flux rope exists prior to the 
CME, plays an essential role in triggering it \citep{fan_05}, and also explains a range of pre-CME observations 
of prominences, cavities, and sigmoids \citep{gibfan_06b,gibetal_07a}, as well as post-eruption phenomena such 
as coronal dimmings and the structure of magnetic clouds \citep{gibfan_08}. However, as it erupts, it 
reconnects internally and with surrounding fields so that it breaks in two, with one portion of the rope 
remaining behind, and the other escaping as the CME and magnetic cloud.

It is very common for a significant portion of prominence mass to remain behind after an eruption 
\citep[see e.g.,][]{gilbert_00}. Because prominence mass is often modeled as being situated within the dips of 
a magnetic flux rope, one possibility is that such 'partially-erupting prominences' occur because of a 
flux-rope bifurcation as predicted by the model. It is often difficult to tell, however, whether 
internal reconnections are indeed occurring, or whether the apparent split in the prominence mass arises from 
the differing evolution of adjacent, but magnetically-disconnected structures. It is therefore essential to consider 
a range of multi-wavelength observations in order to look for additional evidence of internal reconnection 
consistent with predicted model observables. It is worth emphasizing here that it is almost impossible 
to observe reconnection directly in the corona as there is no direct measurement of magnetic field in the corona. 
Therefore, we have to rely on indirect evidences derived from the multi-wavelength observations.

The rest of the paper is structured as follows. In Section 2 we will describe the observables predicted by the 
model. In Section 3 we will present analyses of cases of partially erupting prominences, and consider how well 
the data supports the model. In Section 4 we present our conclusions.

\begin{figure*}
\centering
\includegraphics[width=0.85\textwidth]{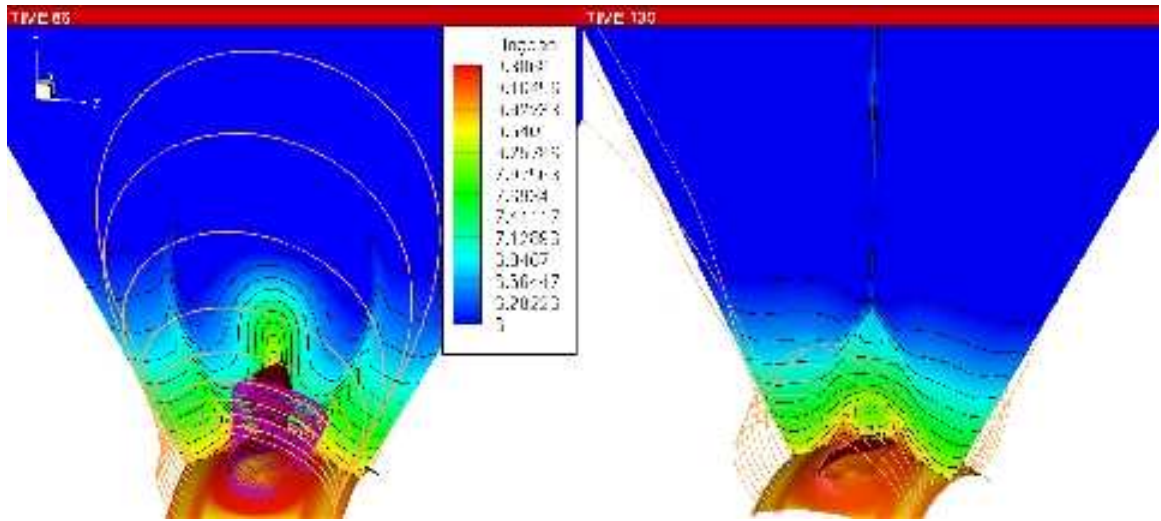}
\caption{PEFR-model partially-erupting prominence and reforming cavity.  Isosurface and isocontours show the logarithm of number density (in cgs units), magenta lines show the bald-patch-separatrix-surface (BPSS) (see text and Figure 2 of \cite{gibfan_06b}), and yellow lines show overlying arcade field.  Evolving, initially-dipped field identified with the prominence is shown in brown.}\label{figcav}
\end{figure*}
\begin{figure*}
\centering
\includegraphics[width=0.85\textwidth]{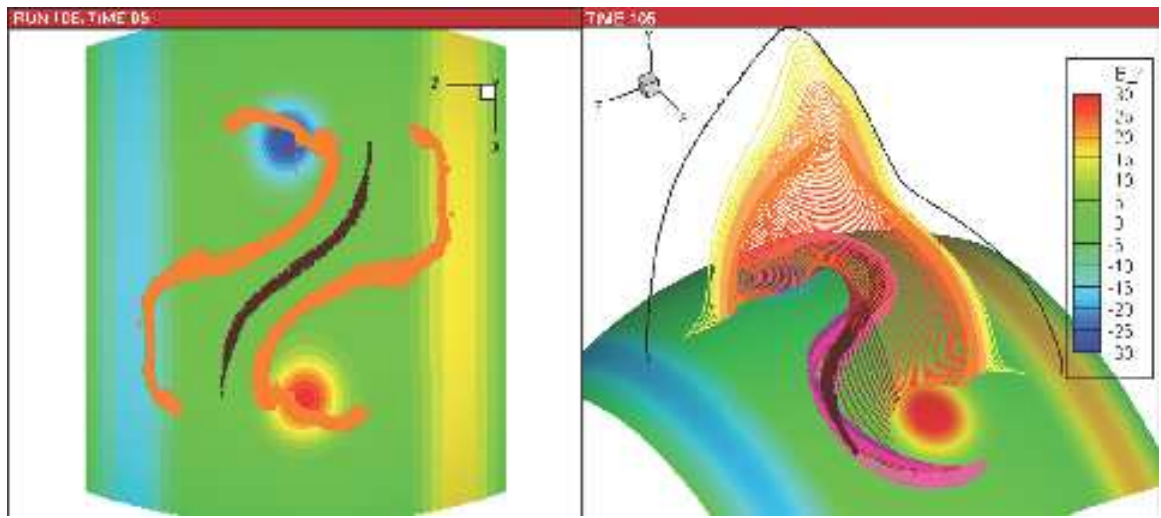}
\caption{Left: Footpoints of reconnected field lines (orange) forming two ribbons surrounding surviving portion of filament (brown). See \cite{gibfan_06a} for discussion of how reconnected field lines are determined. Right: From \cite{gibfan_06b} 
Post-eruption state: cusp over sigmoid and filament. Red-orange-black field lines are sample lines straddling the magnetic neutral line lines show bald-patch-separatrix-surface (BPSS) of surviving rope, brown shows surviving portion of filament.}\label{figribfilsigcusp}
\end{figure*}

\section {{\bf Bifurcating magnetic flux ropes: PEFR model-predicted observables} \label{third_way}}

The concept of a flux rope breaking in two during eruption was first discussed by \cite{gilbert_00}, as 
a means of explaining observations where only a portion of the pre-eruption prominence escaped in 
the eruption (note that  a filament and prominence are the same entity, differing only in how they 
are observed: we will use the terms interchangeably from here on). These authors found that the 
majority of erupting prominences studied demonstrated a separation of escaping material from the bulk 
of the prominence, and proposed that reconnections occurred within the prominence at an X-type 
neutral line which formed during the eruption. In a numerical simulation of the dynamic emergence of 
a magnetic flux rope across the photosphere, \cite{manch_04} demonstrated that shearing motions induced 
by axial field gradients could lead to internal reconnections, and ultimately the rope's bifurcation 
and the upward expansion of its upper portion. \cite{birn_06} likewise demonstrated the formation of 
a current sheet within an unstable flux rope, which separated an outwardly expanding portion of the 
rope from a portion that remained below. \cite{gibfan_06a} described the full evolution of such 
a partially-expelled flux rope (PEFR), from the rope's emergence and formation as a 
pre-eruption equilibrium, through its destabilization, eruption, and bifurcation, and ultimately to 
an end-state with magnetic field closing down over the surviving portion of the rope. This is the 
PEFR model we will specifically consider in this paper. We now summarize the model predictions 
for observables that can be directly compared to data (see \cite{gibfan_06b,gibfan_08} 
and \cite{gibetal_07a} for further details).

The first set of observables provide evidence for a partial eruption of prominence mass:

\begin{itemize}

\item {\bf Ejected prominence mass:} Figure~\ref{figcav} shows the evolution of initially dipped 
field (brown) during the rope's eruption, which we identify with the prominence mass.  Current sheets 
form within the rope, splitting the erupting material in two (visible in the right-hand image as 
the central thin, vertical density enhancement above the reformed cavity).  The upper-most 
material escapes upwards, and is the core of the the three-part (dome/cavity/core) structure of the 
CME.  See also Figures 5 and 8 of \cite{gibfan_06b}.  

\item {\bf Surviving prominence mass:} In Figure~\ref{figcav} right-hand image, the brown material 
lying below the central current sheet is essentially unaffected by the eruption.  Thus, some portion 
of prominence mass is not ejected.

\end{itemize}

\begin{figure*}
\centering
\includegraphics[width=0.8\textwidth]{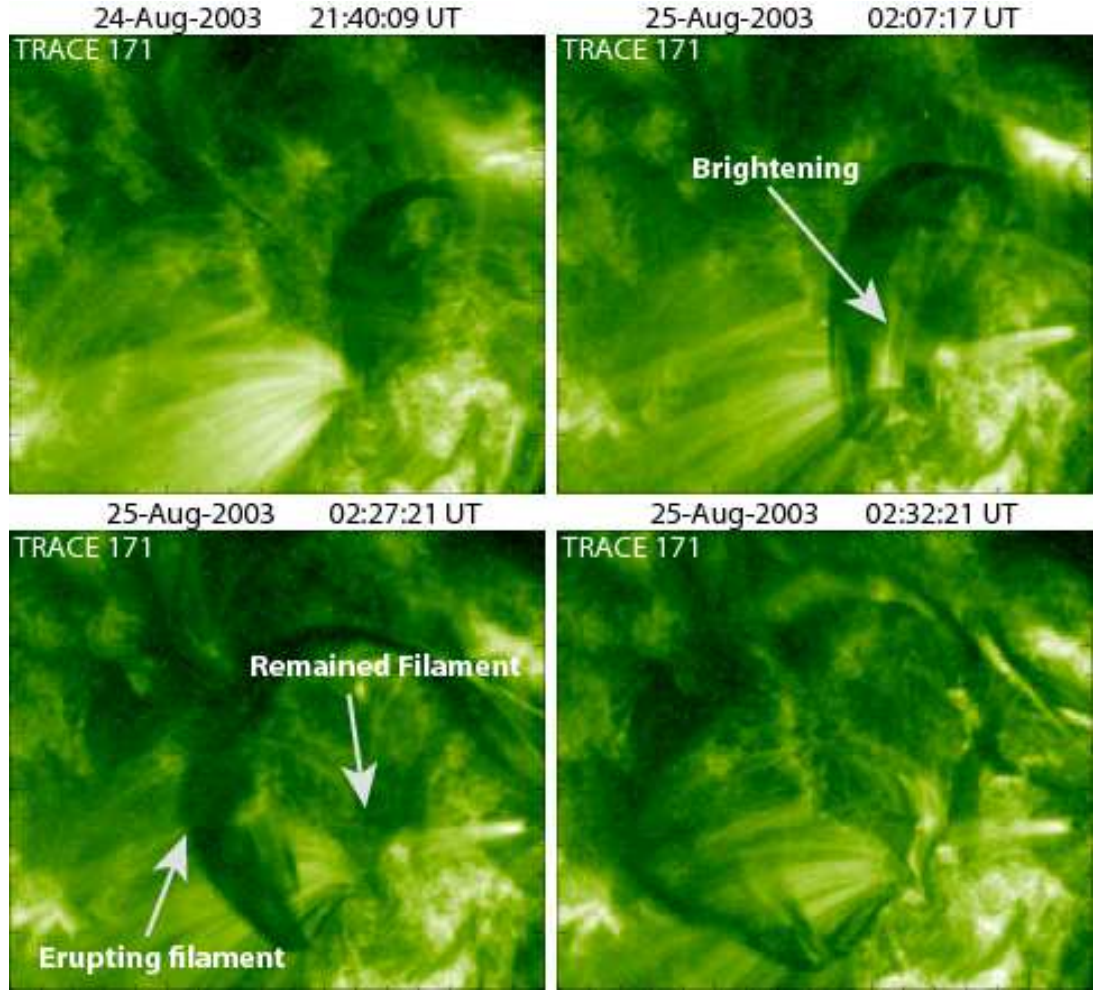}
\caption{TRACE observation of a filament breaking while eruption on 25-Aug.-2003. The brightening, erupting and remaining filament are marked with arrows and labelled. \label{part_def_trace}}
\end{figure*}

Because adjacent, magnetically disconnected structures might create the illusion of partial eruption, 
it is important to consider the next two sets of observables, which provide evidence for 
internal reconnection:

\begin{itemize}

\item {\bf X-type flows:} The prominence-tracing material splits in two because of 
reconnection at the central current sheet (see Fig. 5 in \cite{gibfan_06b}). The model therefore predicts 
mass flows diverging from a central point, so that the upper material would continuously move out, while 
material below the reconnection point might first surge up, but then fall back down.  

\item {\bf Two ribbon flares surrounding non-erupting filament:} Figure \ref{figribfilsigcusp} 
(left panel) shows the footpoints of field lines that have reconnected at the current sheets, with 
the surviving filament shown in brown.  Thus, a two-ribbon flare would bracket the surviving portion 
of the filament, and indicate reconnection above it. Since this signature is also consistent 
with completely non-erupting filament e.g., in case of confined flares, it is important to combine it 
with evidence for ejected filament mass as described above.

\end{itemize}

Internal reconnection within sheared (but not flux-rope) precursor fields could explain all of 
the observables listed so far (see e.g., \cite{tokbell_02}). The observables we now list, however, 
are unique to the PEFR model:

\begin{figure}
\centering
\includegraphics[width=0.4\textwidth]{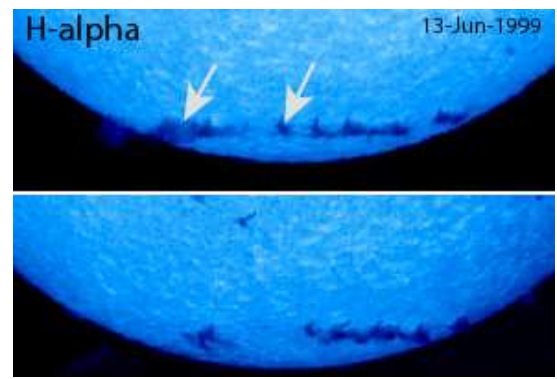}
\caption{BBSO H$\alpha$ Observation of an erupting polar crown filament on 13-Jun-1999. The part of the filament which erupts is marked by an arrow.\label{part_def_halpha}}
\end{figure}

\begin{itemize}

\item {\bf Transition from sigmoid to cusp-overlying-sigmoid state:}  The transition from sheared to 
cusp-shaped arises as reconnections occur initially at the sigmoidal surface separating rope and 
arcade, then on sigmoidal lines within the core of the rope, and finally behind the erupting portion 
of the rope forming the cusp (see bottom panel of Fig. 5 in \cite{gibfan_06b}). Such a transition 
would also occur for non-flux-rope-precursor models, but the PEFR model makes the additional 
prediction that a sigmoid would reform below the cusp after the eruption. This is shown in the right 
panel of Figure~\ref{figribfilsigcusp}, where the surviving portion of the flux rope is illustrated by 
the magenta field lines which represent the critical 'bald-patch-separatrix-surface' (BPSS) of 
dipped field just grazing the ``photosphere'' (i.e., the simulation's lower boundary). The BPSS 
arises from the flux rope topology, and has been demonstrated to be a site where current sheets form 
under perturbation -- not just during eruption -- and so may explain ``quiescent''
(non-eruptive) sigmoids.  The predicted observable of the PEFR model is thus a sigmoid transitioning to 
a cusp which overlays a quickly reforming quiescent sigmoid.

\begin{figure*}
\centering
\includegraphics[width=0.8\textwidth]{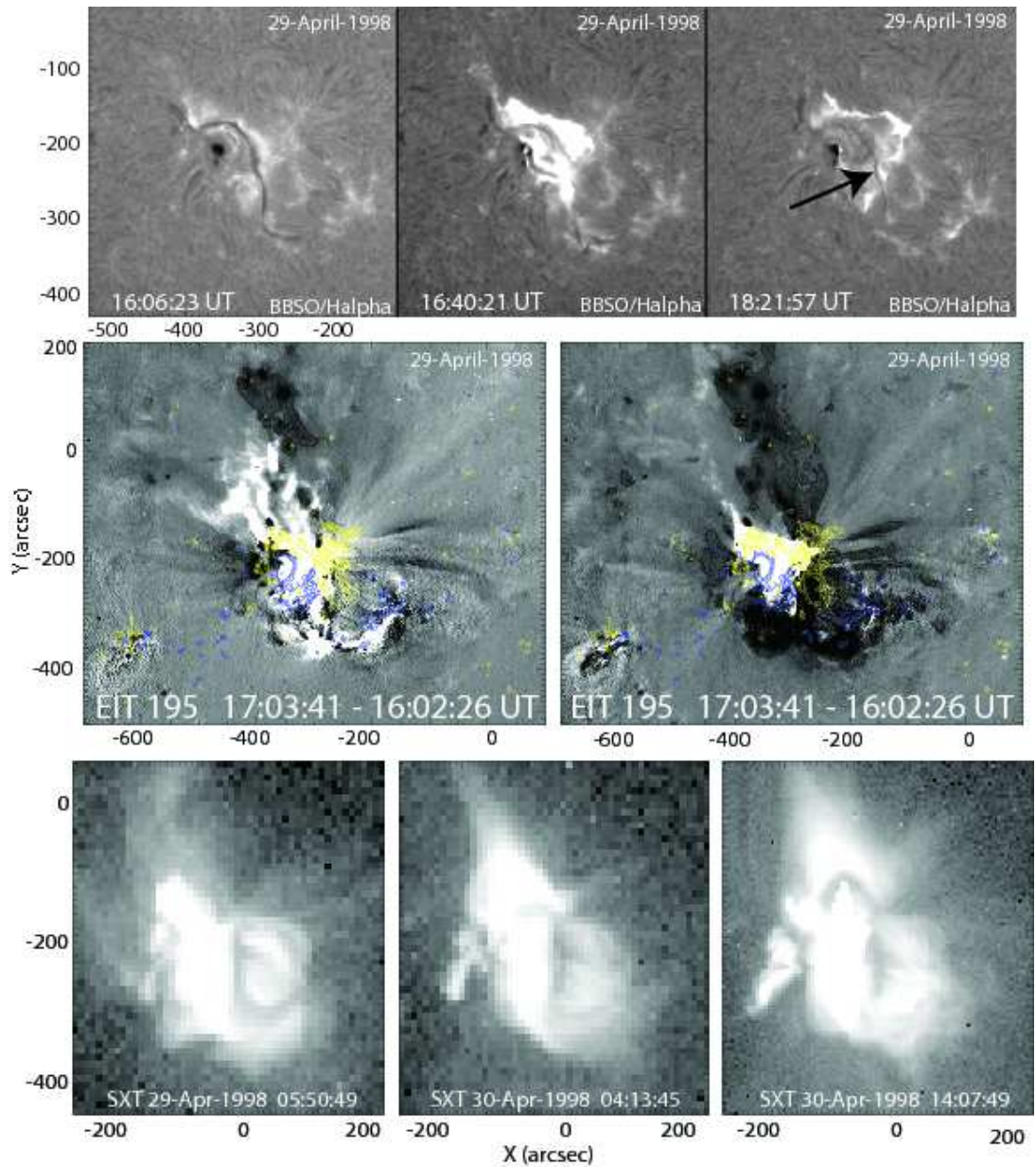}
\caption{Top Panel: H${\alpha}$ images taken from BBSO, showing the erupting filament on 29-Apr.-1998. The arrow in the last image locates
the surviving filament in between two ribbons. Middle Panel: Base difference images taken by EIT at 195~{\AA} on 29-Apr.-1998. Note that
the image recorded by EIT at 16:02~UT was considered as the base image and the images were differentially rotated to line up with the base
image prior to the subtraction. Overplotted black contours show the dimming regions (indicating a 40\% decrease in the intensity relative
to the base image). Yellow contours are positive polarity and blue contours represent the negative polarity regions as was observed by
the MDI magnetograms. Bottom Panel: Yohkoh/SXT images showing the evolution of the source region before and after the eruption, on
29-Apr.-1998 (left image) and on 30-Apr.-1998 (middle and right images).\label{apr_29}}
\end{figure*}

\begin{figure*}
\centering
\includegraphics[width=0.8\textwidth]{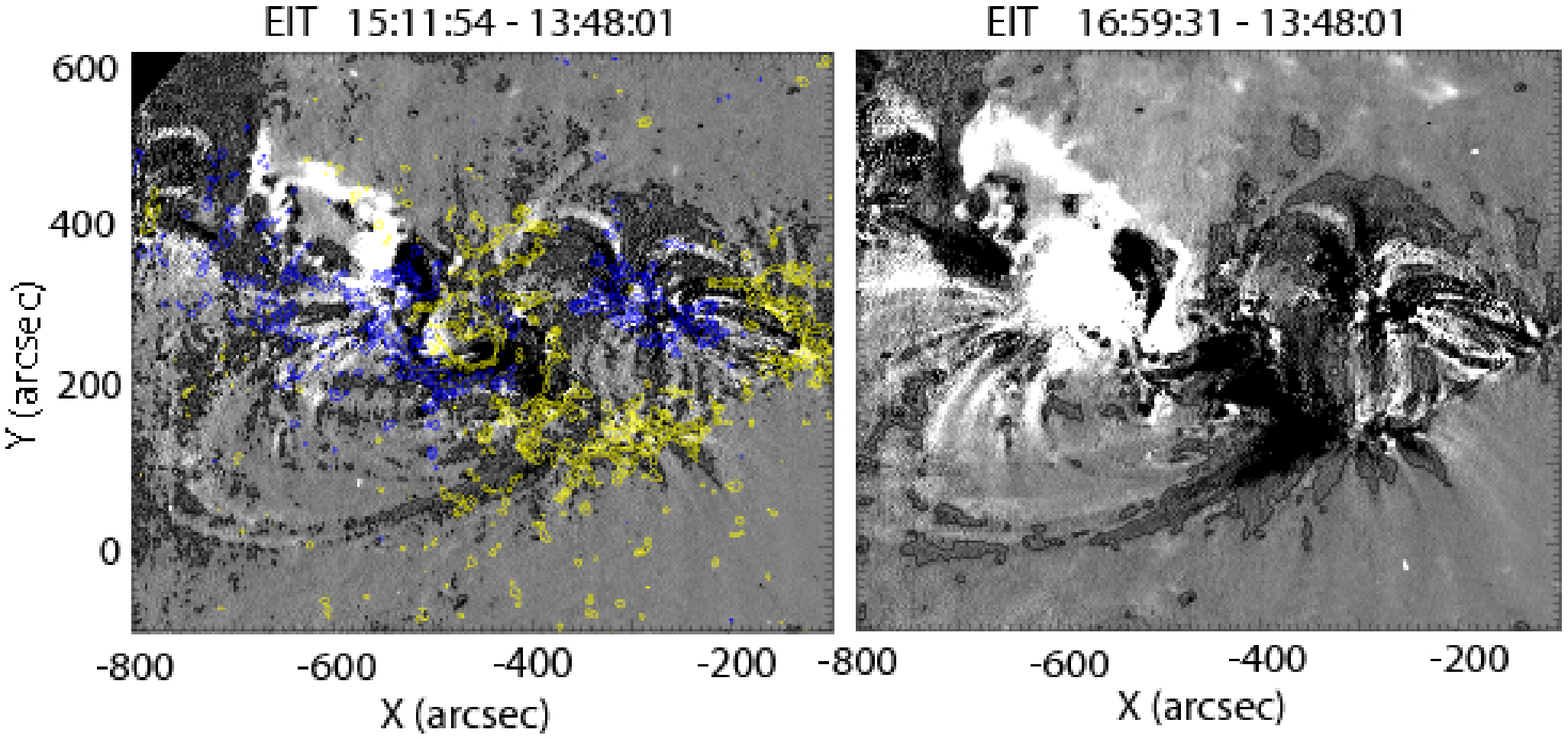}
\caption{Base difference images taken by EIT at 195~{\AA} on 17-Aug.-1999 showing dimming regions overplotted with MDI magnetic
contours, displayed as in Figure \ref{apr_29}.\label{aug_17_eit}.  For this event the base image was taken at 13:48 UT (before the flare).}
\end{figure*}

\item {\bf Transient coronal holes rooted outside original source region:} Another type of 
reconnection explicitly predicted by the PEFR model is connectivity-changing reconnections between 
the rope and the arcade (see bottom panel of Fig. 5 in \cite{gibfan_06b}). These 
``mixed-connectivity'' field lines possess higher, erupting counterparts, which undergo 
further 'rope-breaking' reconnections at the central, vertical sheets to result in an escaping rope 
which is rooted in the original arcade boundary. If transient coronal holes correspond to the 
footpoints of the escaping flux rope, the PEFR model predicts that they would lie completely outside 
the original source region \citep[see Figure 8 in][]{gibfan_08}.

\item {\bf Pre-existing and reforming cavity (subject to line-of-sight visibility):} Note that the PEFR model 
also predicts that both the initial, and the surviving filaments are contained within a region of decreased 
density, i.e. a cavity (Figure \ref{figcav}).  The cavity is an integral part of a flux rope model, with 
a sharply-defined circular boundary arising from a magnetic flux surface \cite{low96,low99}. Thus, the survival 
of the lower flux rope would predict a reformation of the cavity after the eruption. Note, however, that such 
a reforming cavity would only be likely to be observed for partial eruptions of near-limb, large-scale-prominences
without significant intervening structures.  Although partially-erupting filament/cavity systems have been reported
\citep[e.g.][]{riu}, in such cases it is particularly difficult to rule out the eruption of adjacent 
structures along the line of sight. Because the observables providing evidence of internal reconnection tend to 
be best viewed on-disk, the events studied in this paper do not allow us to look for evidence of a reforming cavity.

\end{itemize}

\section{{\bf Observations and analysis}}\label{obs}

The model predicts partially-erupting filaments, so the first test of
its plausibility is to consider how common they may be. We find as
\cite{gilbert_00} did, that some sort of splitting of prominence
material occurs in many, if not most, cases of eruption. In some
cases, a prominence is seen at a viewing angle from which it appears
to rise as a whole, and subsequently breaks in two with respect to its
height, e.g. Fig.~\ref{part_def_trace}. Such cases, particularly when
coupled with brightenings and/or diverging flows at the break points,
are convincing examples of filaments breaking at an internal
reconnection point (see, e.g., \cite{trip_inflow06, trip_inflow07}).
However, other eruptions happen along the length of a filament, e.g.,
Fig.~\ref{part_def_halpha}. Although these too could be due to
internal reconnection, one can not generally rule out the possibility
that the eruption separated two structures that were not
magnetically-connected to begin with.

It is also important to pay close attention to time scales of the
filament's dynamic evolution.  Two-thirds of erupting filaments reform
in the same place and with much the same shape within 1 to 7 days
\citep{priest_book}.  Moreover, filaments are likely to be heated
during eruption so that they may temporarily leave the H$\alpha$
bandpass.  If so, they might be better seen in EUV or SXR
observations. Hence, multi-wavelength observations are important in
establishing the timing and extent of filament eruptions.

\begin{figure*}
\centering
\includegraphics[width=0.8\textwidth]{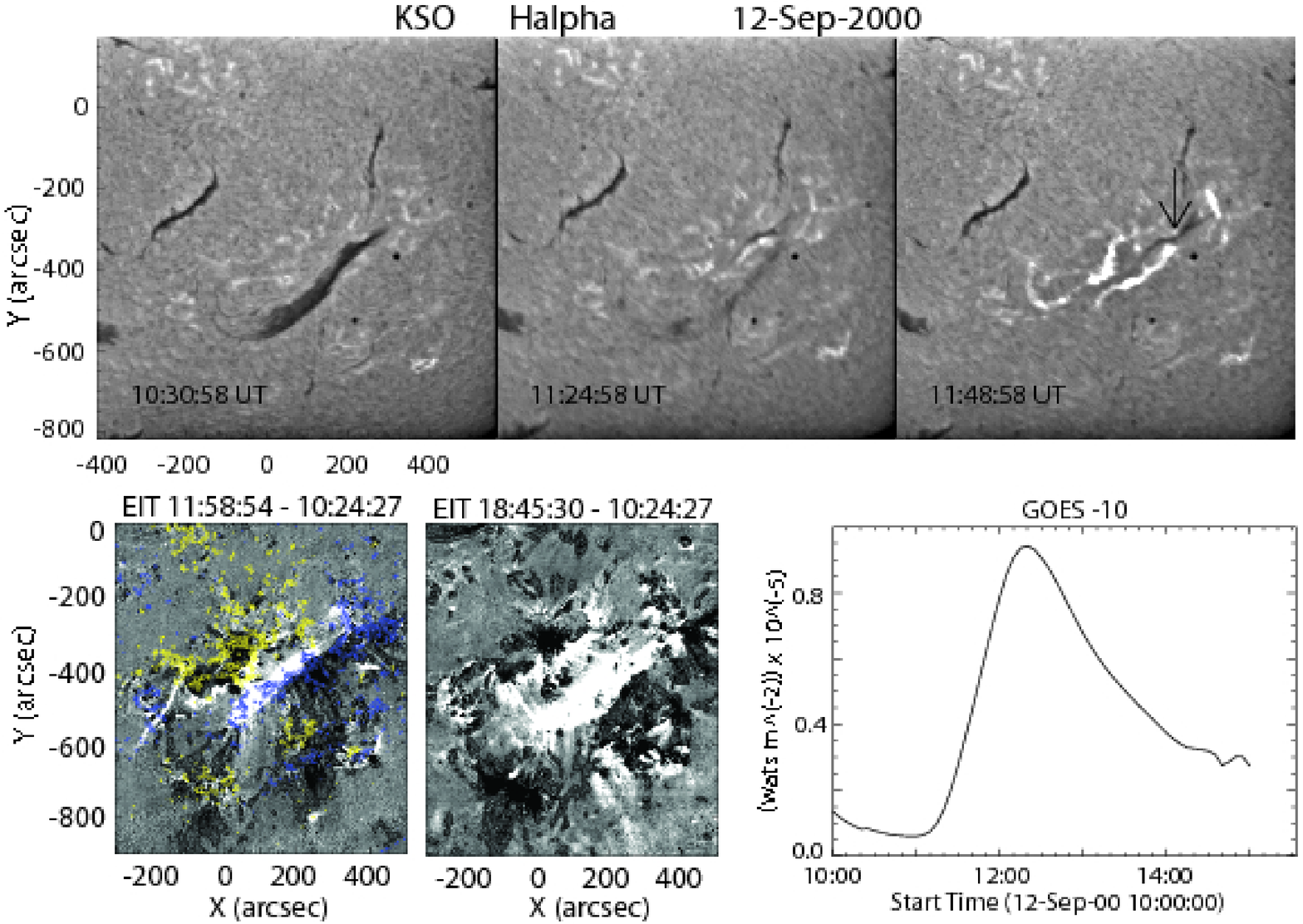}
\caption{Top panels: BBSO H${\alpha}$ images showing the partially erupting filament on 12-Sept-00. The erupting and surviving filament
is marked by arrows and labelled accordingly. Bottom panel: EIT 195~{\AA} base difference images overplotted with MDI magnetic field
contours, displayed as in Figure \ref{apr_29}.\label{sep_12}. For this event the base image was taken at 10:24 UT (before the flare).}
\end{figure*}

\begin{figure*}
\centering
\includegraphics[width=0.8\textwidth]{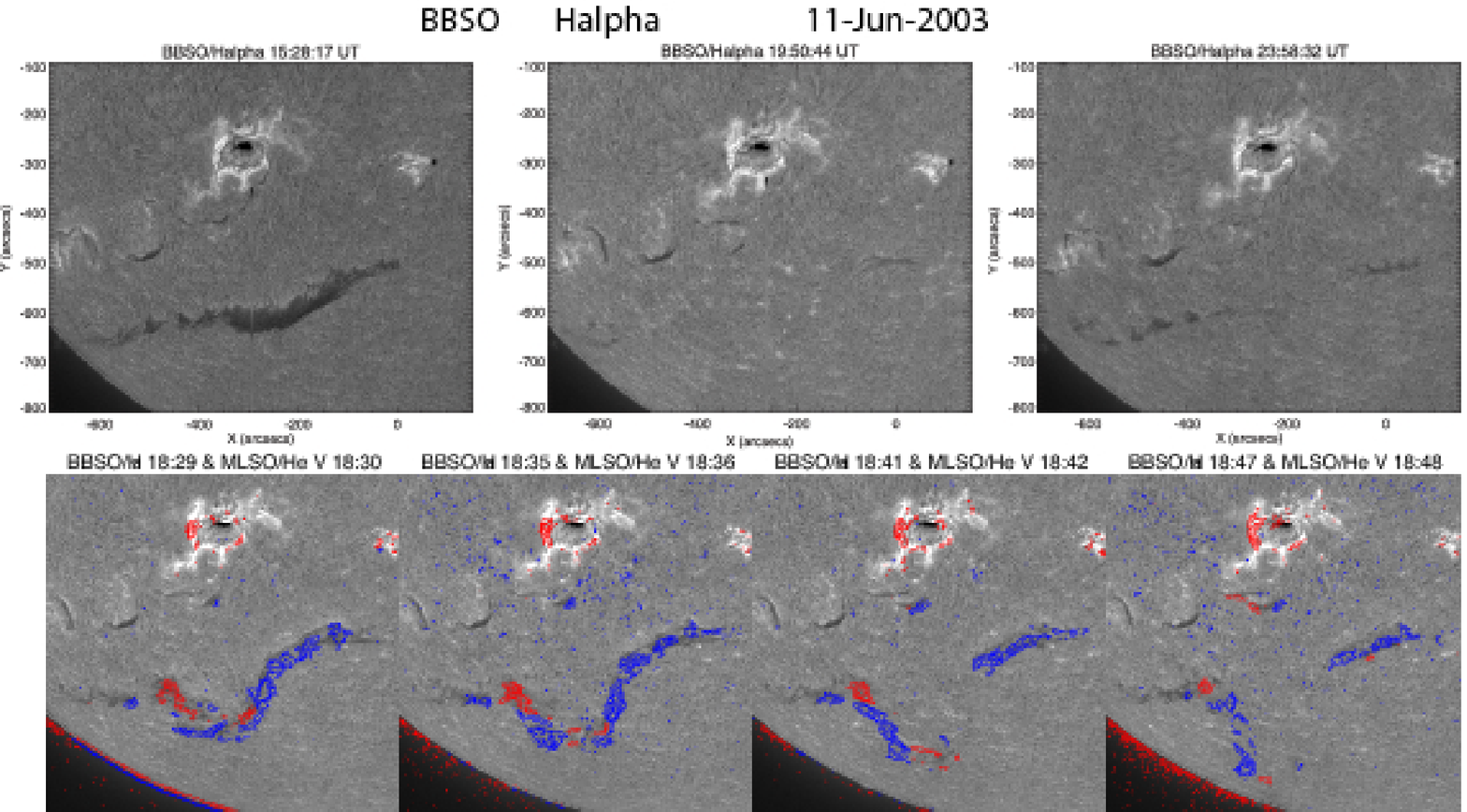}
\caption{Top panels:H${\alpha}$ images recorded at BBSO showing the erupting filament (left panel) and filament reformation (right panel) on 11-Jun.-2003. 2nd, 3rd and 4th rows: BBSO H-alpha images overlaid by contours of MLSO/CHIP velocity data on 11-Jun.-2003. Red (blue) contours indicate motions away from (toward) the observer. \label{jun_11}}
\end{figure*}

In choosing our partial-eruption cases, therefore, we require the following:

\begin{enumerate}
\item {\bf Ejection of filament material}, in particular the presence of a core within the associated CME
\item {\bf Survival of material}, in particular the reappearance of a filament in H$\alpha$ and/or EUV
\item{\bf Evidence for internal reconnection}, as described in the model observables above.  
\end{enumerate}

\begin{figure}
\centering
\includegraphics[width=0.45\textwidth]{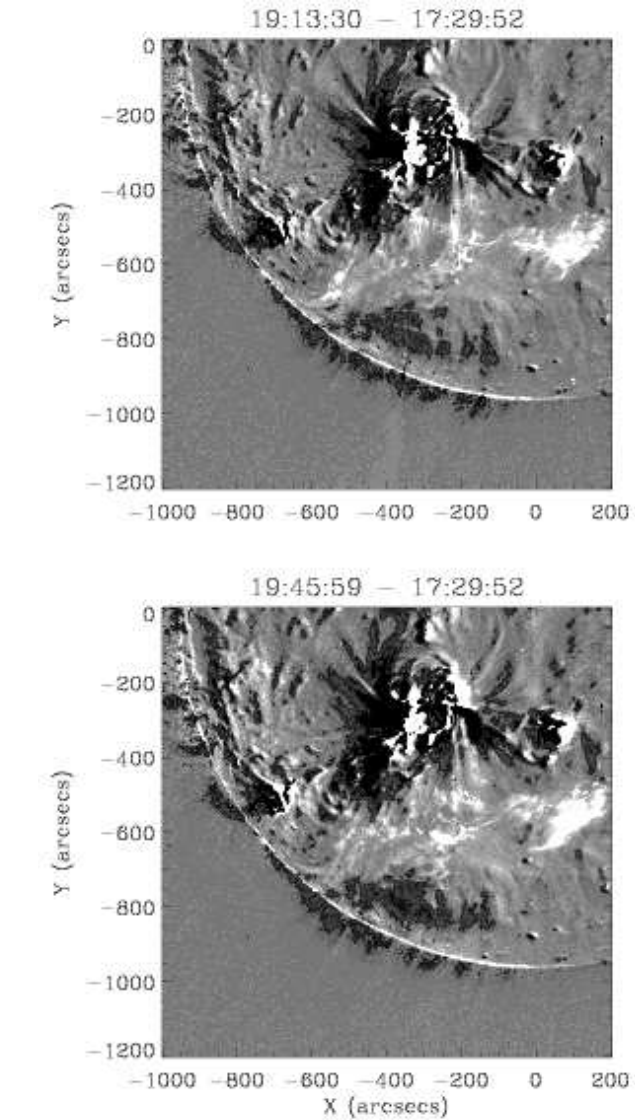}
\caption{Base difference images taken by EIT at 195~{\AA} on 11-Jun.-2003.\label{jun11_eit}}
\end{figure}

\subsection{Event on April 29, 1998}

An erupting filament associated with an M-class flare was observed on
29-Apr.-1998. The eruption was also observed by the BBSO H${\alpha}$
telescope (top row images in Fig.~\ref{apr_29}), the EIT at 195~{\AA}
(middle row in Fig~\ref{apr_29}), and by the SXT (bottom row in
Fig~\ref{apr_29}).

{\bf Evidence for both ejected and surviving material:} The filament
eruption was associated with a halo CME as observed by the LASCO/C2
coronagraph at 16:58:54~UT. Although the CME was a halo, a white-light
core is visible in the structure which can be associated with the
erupting filament, that was also visible as ejected material in EIT
running difference observations.  The top panel in Fig.~\ref{apr_29}
displays images of the erupting filament and associated flare as
observed in H${\alpha}$.  A dark filamentary structure, marked with an
arrow in the last image, is located at the same position as that of
the original filament before the eruption, providing evidence of
surviving material.

{\bf Evidence for internal reconnection (non-PEFR unique):} The
chromospheric counterparts of the flare are two ribbons, seen in the
top panel images with the dark filament in between. The middle image
is recorded at around the peak of the flare. As the ribbons separate
outwards from the neutral line, the contrast of the filamentary
structure against the flare emission improves and it is seen more
clearly (see the last image).

{\bf Evidence for observables unique to PEFR model:} The bottom panel
shows SXR loops: The left image shows a clear sigmoidal system of hot
loops overlying the sigmoidal filament seen in the H${\alpha}$
observations. After the eruption the sigmoidal region has transitioned
to cusp-shaped post-eruption arcades (PEAs) overlying a sheared system
of loops (see the middle and last images in the bottom panel of
Fig.~\ref{apr_29}). The images in the middle row of Fig.~\ref{apr_29}
displays EIT 195~{\AA} base difference images (in background) in which
a fixed image frame was subtracted from the following images. Base
difference images provide information about overall changes in the
source region with respect to a fixed image frame.  In this case, they
demonstrate that the dimming regions lie outside the eruption source
region.

\subsection{Event on August 17, 1999}

An erupting filament, associated with a C-class flare was observed on
 17-Aug.-1999 (Flare start: 14:37~UT; peak: 15:33~UT).
  
 {\bf Evidence for both ejected and surviving material:} The filament
 eruption was associated with a CME with a bright core as observed by
 LASCO/C2 at 15:30:05~UT. The H$\alpha$ data recorded by BBSO showed a surviving filament about 2.5 hours
 post-flare (see Fig. 1, 2nd row, 2nd column image in \cite{gibetal_02}).
 
 {\bf Evidence for internal reconnection (non-PEFR unique):} The
 surviving filament is clearly straddled by two flare ribbons (also
 seen towards the north, labelled and marked by arrows in Fig. 1, 2nd row, 2nd column image 
in \cite{gibetal_02}). The filament
 actually became more pronounced during the eruption, and continued to
 develop and grow with time (see Fig. 1, 3rd row, 2nd column in \cite{gibetal_02}). 
 
 {\bf Evidence for observables unique to PEFR model:} The SXT observations recorded a day before eruption 
showed a clear sigmoidal structure. After the eruption, a cusp-shaped structure lying over a part of 
the surviving sigmoid was seen (see SXT images in Fig. 1 in 
\cite{gibetal_02}). Fig~\ref{aug_17_eit} displays base difference images taken by the EIT, demonstrating 
that the dimming regions predominantly occur outside the main eruption location (see the right panel).

For further details on this event, see \cite{gibetal_02}.

\subsection{Event on September 12, 2000}

A spectacular quiescent filament eruption was observed on 12-Sep.-2000
associated with an M-class flare (start: 11:06~UT; peak:11:55~UT).

 {\bf Evidence for both ejected and surviving material:} The eruption
was also associated with a three-part structured CME as observed by
LASCO/C2 at 11:54~UT. The top panel in Fig.~\ref{sep_12} displays the
erupting filament detected by the H$\alpha$ telescope at Kanzelhoehe
Solar Observatory.  In the middle image, taken at 11:24 approximately
20 mins after the eruption, when the erupting filament is already at
some height, a dark filament towards the northwestern side in the
source region can be seen. The erupting and the surviving filaments
are marked and labelled in the top middle image. This provides
evidence that the filament breaks in the middle towards its
north-western end during eruption.

{\bf Evidence for internal reconnection (non-PEFR unique):} The
surviving filament is clearly straddled by two flare ribbons (last
image in top panel of Fig.~\ref{sep_12}).

 {\bf Evidence for observables unique to PEFR model:} The bottom panel
of Fig.~\ref{sep_12} displays the base difference images taken by the
EIT at 195~{\AA} overplotted with magnetic field contours. As can be
depicted from the figure, the dimming is seen not at the foot points
of the erupting filament but in the surrounding region. SXR
observations were not available for this event (or others in this
paper that post-dated Yohkoh observations and pre-dated Hinode), but
the second image in the bottom panel shows the formation of
post-eruption arcades which appear to be more sheared towards the
north above the surviving filament.

\subsection{Event on June 11, 2003}

An eruption of a filament was observed on 11-Jun-2003 near the eastern
limb. 

{\bf Evidence for both ejected and surviving material:} The top panel
in Fig.~\ref{jun_11} shows three BBSO/H-alpha images before (left
image), during (middle image)and after the eruption (last image). The
eastern and western legs of the polar crown filament reform quickly,
but its middle part never reforms.  A CME with a three-part structure
was observed by the MLSO Mk4 coronameter.

{\bf Evidence for internal reconnection (non-PEFR unique):}  
Filament
material is observed to flow back to the Sun's surface during the
eruption in both H$\alpha$ and EIT movies. 
The bottom three panels in Fig.~\ref{jun_11} show BBSO H$\alpha$
images overlaid with contours of MLSO/CHIP velocity data. Red (blue)
contours indicate motions away from (toward) the observer. During
eruption, the line-of-sight velocity reveals that plasma is moving
both towards and away from the Sun being predominantly away, showing
that the prominence is moving upward. In the early rise phase of the
prominence plasma motion in both
directions is evident at the middle of the prominence. Later on, the
plasma motion towards the Sun is more dominant towards the eastern leg
of the prominence, probably indicating the draining of plasma along
the legs.

 {\bf Evidence for observables unique to PEFR model:} 
Figure~\ref{jun11_eit} displays base difference images taken by the
EIT at 195~{\AA}. An image frame taken at 17:29:52 before the eruption
was considered as the base image. Since this event was also on the
limb, we did not have MDI magnetic field measurements with good
sensitivity. However, from Fig.~\ref{jun11_eit}, the dimming regions
appear to be outside the erupting source regions.

\subsection{Event on August 25, 2003}

A relatively small erupting filament was observed on August 25, 2003
at around 02:00~UT. 

\begin{figure*}
\centering
\includegraphics[width=0.8\textwidth]{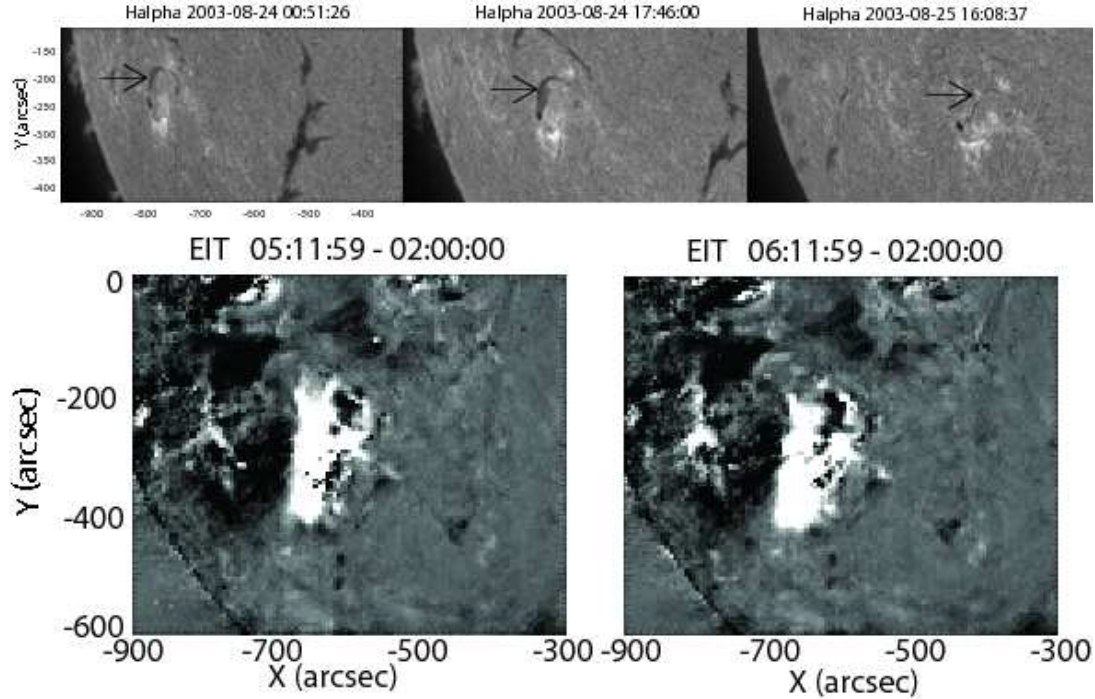}
\caption{Top Panel:H${\alpha}$ images taken from BBSO, showing the filament before (left and the middle panel) and after the eruption
(last panel). Arrows demark the filament which is of interest. Bottom panel: EIT at 195~{\AA} base difference images taken on 25-Aug.-2003
displayed as in Figure \ref{apr_29}. \label{aug_24}. For this event the base image was taken at 02:00 UT (before the flare).}
\end{figure*}

{\bf Evidence for both ejected and surviving material:} The eruption
was associated with a white-light CME comprising a bright core as
detected by the LASCO/C2 at 03:25~UT.  Figure.~\ref{part_def_trace}
displays images taken by the TRACE at 171~{\AA}. The filament starts
with a slow rise phase at around 23:00~UT. At 02:07~UT (top right
image in Fig.~\ref{part_def_trace}), when the filament has risen some
height, it appears to separate in two, leaving some filament material
behind.  The erupting and surviving filaments are marked with arrows
and labelled in the bottom left panel.  The top panel (left and middle
images) of Fig.~\ref{aug_24} display the BBSO H$\alpha$ filament a day
before eruption (marked with arrows).  Although surviving material is
evident in the TRACE images even during eruption, it was not seen in
$H{\alpha}$ until the next day (top right image of Fig.~\ref{aug_24},
marked with an arrow). This may be due to heating during eruption, and
emphasizes the importance of multi-wavelength observations when
identifying the partial eruptions.

{\bf Evidence for internal reconnection (non-PEFR unique):} A
brightening is seen in between the erupting part and the surviving
part marked in the top right panel of Fig.~\ref{part_def_trace}. This
brightening may indicate energy release due to reconnections within
the erupting filament.  

{\bf Evidence for observables unique to PEFR model:} The bottom panel
of Fig.~\ref{aug_24} displays EIT 195~{\AA} base difference images.
Since this event occurs far from the disk center, we did not have
enough sensitivity for magnetic field data. However, the EIT dimming
observations on their own demonstrate that the dimmings are
predominantly outside the source region of the eruption.

\subsection{Event on May 13, 2005}

An erupting filament was observed on 13-May-2005 associated with an
M-class flare (start:16:23~UT, peak:16:53~UT). 

\begin{figure*}
\centering
\includegraphics[width=0.8\textwidth]{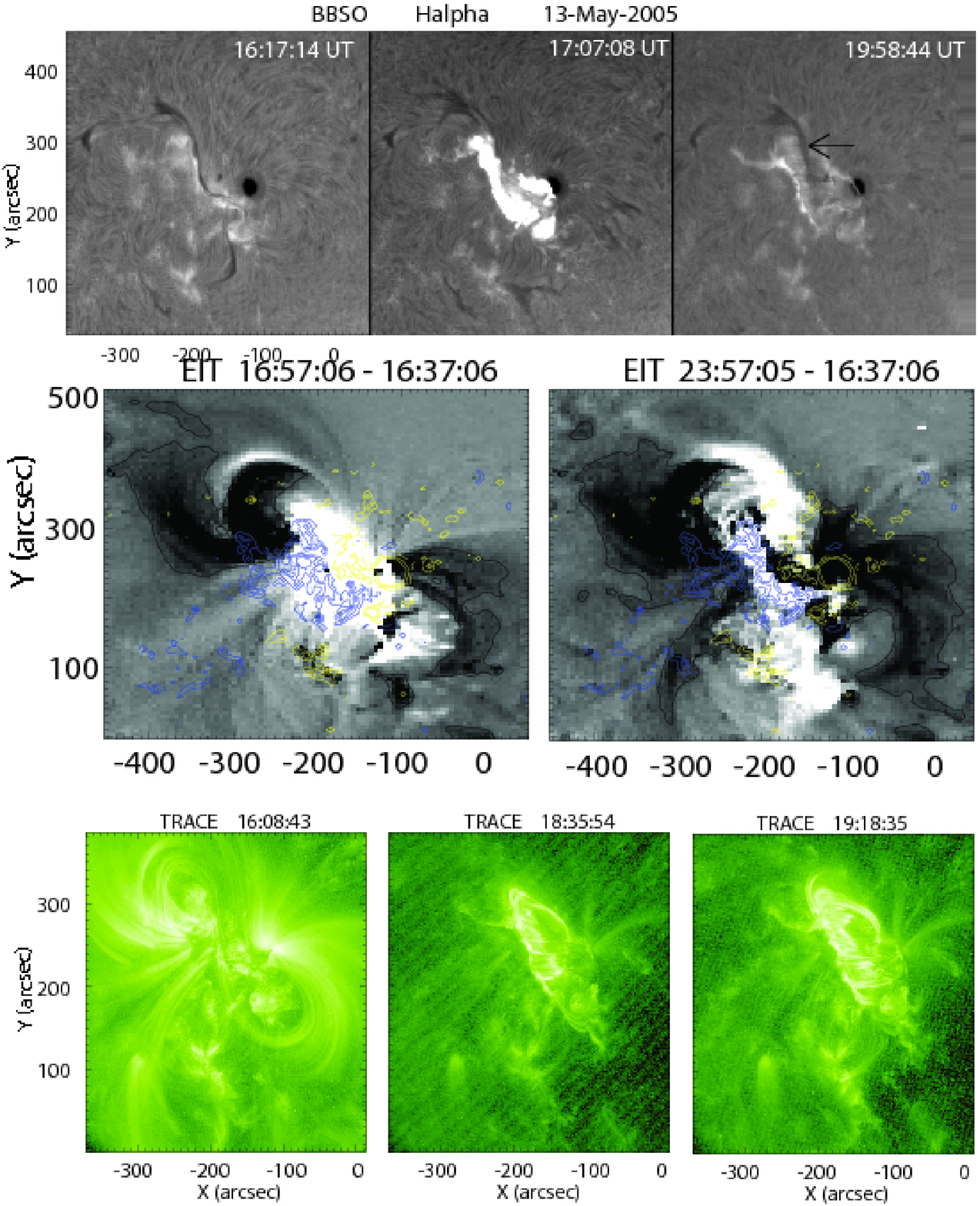}
\caption{Top panels: BBSO H${\alpha}$ images showing the filament before eruption (left panel), associated two ribbon flare during eruption (middle panel) and surviving filament in between two ribbons (right panel). The arrow in the last frame locates the surviving filament. Middle panels: Base difference images taken by EIT at 195~{\AA} on 13-May.-2005, overplotted with MDI magnetic contours, displayed as in Figure \ref{apr_29}.  For this event the base image was taken at 16:37 UT (before the flare). Bottom panels: Images taken by TRACE on 13-May-2005 at 171~{\AA} before (left panel) and after (middle and right panel) the eruption. \label{may_13}}
\end{figure*}

{\bf Evidence for both ejected and surviving material:} The filament
eruption was associated with a CME comprising a bright core as
detected with LASCO/C2. The top panel in Fig.~\ref{may_13} displays
the H${\alpha}$ images taken from BBSO, showing the filament before
eruption (left panel), and the associated flare (middle panel). The
right image recorded about 4 hours after the eruption clearly shows
the surviving filament (marked with an arrow).

{\bf Evidence for internal reconnection (non-PEFR unique):} The right
image also shows that a two-ribbon flare brackets the surviving
filament.

 {\bf Evidence for observables unique to PEFR model:} The middle panel
 images in Fig.~\ref{may_13} display EIT 195~{\AA} base difference
 images overplotted with magnetic field contours on top. As is evident
 from the figure, the dimming regions are outside the source region of
 eruption and the area of dimming region increases with time and
 expands outwards.  Although no SXR observations were available, the
 bottom panel in Fig.~\ref{may_13} displays TRACE observations of the
 eruption. A highly sheared pre-eruption sigmoidal region was visible
 before the flare (left image, bottom panel).  After the eruption
 (right image) there remain loops that are more sheared towards the
 north (where the surviving part of the filament resides, see right
 image in the top panel) than in the southern part of the flaring
 region.

\begin{table*}
\caption{Comparison between the PEFR model observables described in section~\ref{third_way} and observations.
\label{summary_table}}
\begin{center}
\begin{tabular}{lcccc}
\hline
\hline

						& {\bf Evidence of}     			& {\bf Evidence of}     		    & {\bf Evidence of}            		& {\bf Evidence of}      		\\
                       	& {\bf ejected}       				& {\bf surviving}  	     			& {\bf internal reconnections} 		& {\bf internal reconnections}   \\
                       	& {\bf filament  material} 			& {\bf filament material} 	     	& {\bf (not unique to}  	       	& {\bf (unique to} 				 \\
           				&		 			 				&    	     						& {\bf PEFR model)}   				& {\bf PEFR model)}  			\\
\hline
{\bf 29-Apr.-1998} 	    & white-light CME core;             & reformed filament           		& flare ribbons                    	&sigmoid to\\
                       	& ejected material visible    		& visible within   					& straddling surviving     			&cusp-over-sigmoid (SXR);\\
                       	& in EIT running difference     	& two hours           				& filament        					&dimmings outside source (EUV)\\
\hline

{\bf 17-Aug.-1999}     	& white-light CME core             	& reformed filament           		& flare ribbons                    	&sigmoid to\\
                       	& 			    					& visible within   					& straddling surviving     			&cusp-over-sigmoid (SXR);\\
                       	& 				       				& two and a half hours           	& filament        					&dimmings outside source (EUV)\\
\hline

{\bf 12-Sep.-2000} 	  	& white-light CME core           	& reformed filament           		& flare ribbons                    	&No SXR\\ 
                       	& 			    					& visible within   					& straddling surviving     			&observations available;\\
                       	& 				       				& twenty minutes          			& filament        					&dimmings outside source (EUV)\\
\hline
{\bf 11-Jun.-2003}     	& white-light CME core             	& reformed filament                	&  Simultaneous                 	&No SXR \\
                       	&     & visible within             	& red and blue shift      			&observations available;						\\
                       	&  		& four hours               	& in erupting filament (\ion{He}{ii})&dimmings outside source  							\\
\hline
{\bf 25-Aug.-2003}     	& white-light CME core;         	& surviving filament           		&brightening and diverging       	&No SXR\\
                       	&      								& material in TRACE;  				& flows at apparent break    		&observations available;\\
                       	&        							& reformed next day in H-$\alpha$          		& point (EUV and white-light) 	&dimmings outside source\\
\hline
{\bf 13-May-2005}      	& white-light CME core           	& reformed filament           		& flare ribbons                    	&No SXR: but EUV \\ 
                       	& 			    					& visible within   					& straddling surviving     			&pre-eruption sigmoid\\
                       	& 				       				& four hours          				& filament        					&and sheared loops\\
                        &									&									&									&above surviving filament;\\
                       	&									&									&									&dimmings outside source\\
\hline
\hline
\end{tabular}
\end{center}
\end{table*}

\section {Discussion and conclusions}

In this paper we have considered multiple cases of partially erupting prominences and have studied their relationship 
with other CME-associated phenomena such as CME three-part structure, two-ribbon flares, mass flows during eruption, 
soft X-ray sigmoids and cusps, and coronal dimmings.  In order to test the plausibility of the partially-expelled flux 
rope model of \cite{gibfan_06a}, we have directly compared these observations to predicted PEFR-model 
observables.  Table~\ref{summary_table} shows that all of the events meet our criteria for partial eruption, that is, 
evidence is observed both for ejected and surviving material, and indirectly for internal reconnection. 
Moreover, every event showed one or both of the PEFR-specific predicted dimmings external to source, and sigmoid 
to cusp-over-sigmoid transition. 

One of the main goals in CME science is to achieve a clear
understanding of the pre-CME magnetic field configurations and their
evolution. Most CME-initiation models and space weather predictions
depend on this.  It is therefore worth considering how well our
observations distinguish between model predictions.  Three competing
possibilities, as discussed in Section~\ref{third_way}, are 1) {\it
Model~IS} ({\it in-situ} forming flux ropes), 2) {\it Model~TE} (total
eruption of pre-existing flux rope) and 3) {\it Model~PEFR} (partial
eruption of pre-existing flux rope). Table~\ref{compare_table} shows
the predicted observables of each, and demonstrates the uniqueness of
the PEFR model (however, see \cite{mandrini07} for an alternate
interpretation of dimmings external to source).  This, in combination
with the results of our observational study, is strong evidence of the
plausibility of the PEFR model.  This in turn argues that while a
magnetic flux rope may well be present prior to eruption, magnetic
reconnection appears to be highly significant during eruption.  Such
reconnection goes beyond merely closing down the magnetic field behind
the erupting and expanding flux rope, by playing a crucial role in the
bifurcation of the flux rope.

\begin{table*}
\caption{Comparing the observables predicted by {\it Model~PEFR} (see
Section~\ref{third_way}) with {\it Model~IS} and {\it
Model~TE}. \label{compare_table}}
\begin{center}
\centering
\begin{minipage}{0.9\textwidth}
\centering
\begin{tabular}{cccc}
\hline
\hline
Predicted observables               	& Model IS             			& Model TE            	& Model PEFR \\
(section~\ref{third_way})        		& ({\it In-situ} forming    	& (Total eruption    	& (Partial eruption\\
										& flux rope) 					& of flux rope)     	& of flux rope)\\
\hline
1 (Pre-existing cavity)			       	&              					&$\surd$                & $\surd$ \\

2 (Quiescent sigmoid)			       	&              					&$\surd$                & $\surd$ \\

3 (Partly erupting filament)		  	& $\surd$             			&                    	& $\surd$ \\

4 (Partly erupting cavity)       		&                     			&                    	& $\surd$ \\

5 (Flare ribbon surrounding filament) 	& $\surd$     			        &       	         	& $\surd$ \\

6 (Sheared post-eruption loops)         & $\surd$             			&                    	& $\surd$ \\

7 (Cusp over reformed sigmoid)    		&                     			&                    	& $\surd$ \\

8 (Dimming external to source)   		& Note\footnote{See \cite{mandrini07} for a possible exception.}    &                    & $\surd$ \\
\hline       
\end{tabular}
\end{minipage}
\end{center}
\end{table*}

An interesting avenue of future work would be to consider limb
observations of partially-erupting prominences.  As discussed above,
line-of-sight issues make establishing partial eruption of prominences
at the limb more difficult.  Two-ribbon flares and sigmoids are not
visible, and clearly identifying diverging flows is likewise
complicated because draining of plasma along the leg of the prominence
is an extremely common phenomena. This draining of plasma does not
necessarily mean that the prominence has broken while eruption, but
could merely arise from plasma sliding back down along field lines
that have been pulled radially. The first and rather plausible example
of breaking of a prominence at the limb during its eruption was shown
by \cite{trip_inflow06, trip_inflow07} based on multi-wavelength
observations including the EIT, the LASCO, and MLSO CHIP data.  It
would be worth looking for more such cases, particularly as STEREO
observations are now allowing us to consider cases where we would have
Earth's-view on-disk observations simultaneous with STEREO EUV limb
observations (the SXR observations from the Hinode satellite would
enable sigmoid observations as well).  Such studies would also resolve
line-of-sight ambiguities and so allow a meaningful consideration of
significance of reforming cavities.

Another motivation for considering limb observations in future is that
writhing motions are best observed at the limb.  Such motions are a
PEFR-model observable that we have not yet mentioned, due to the kink
instability that triggers the eruption in that model (see
\citet{fan_05} for discussion).  We did not observe any apparent
rotation of filaments during the eruption of any of the disk events
studied in this paper.  This could be due to the fact that the time
scale of the rotation of these filaments is smaller than the cadence
of our data. In an independent study using TRACE data \cite{green}
studied 7 active region filaments which rotated during eruption. In
their study, four events were failed eruptions. It is possible that
events in which ejection of material occurs expand more quickly, so
that rotation is not visible on the disk.  Rotation of material is
more easily observed in projection at the limb (see e.g. \cite{riu},
and \cite{gibfan_08}).

Finally, the model for partial eruption of the flux rope can have a
substantial significance to space weather predictions. Partial
eruption from a region means that magnetic energy is still stored in
the surviving twisted field, increasing the likelihood of the region
producing homologous flares and CMEs \citep[see e.g.,][]{cheng,
hudson}. Furthermore, the possibility of partial eruption should be
taken into account when studying the geo-effectiveness of CMEs due to
a possibility of significant differences between the magnetic field
orientations and even topologies of erupted flux rope and that
predicted from the CME's source region field configuration
\citep{gibfan_08}.

\section{Acknowledgements}
DT and HEM acknowledge the support of STFC. DT also acknowledges the support from the High 
Altitude Observatory (HAO), and thanks HAO for its hospitality during his visit. The National Center 
for Atmospheric Research is sponsored by the National Science Foundation. This work has beneﬁted 
from discussions within the Coronal Prominence Cavity International Team (Leader, Sarah Gibson) of 
the International Space Science Institute (ISSI), Bern, Switzerland. We thank B. Kliem, B.C. Low, Guiliana de Toma,
Yuhong Fan, and Scott McIntosh for various useful discussions and comments, and Giuliana de Toma 
for internal HAO review of this paper. We thank the HAO, EIT, LASCO, TRACE, SXT, GOES teams for 
providing the data. SoHO is a mission of international collaboration between ESA and NASA. TRACE is 
a mission of the Stanford-Lockheed Institute for Space Research, and part of the NASA Small Explorer
program. We acknowledge the SURF for providing data for use in this publication. We 
acknowledge observations by Global H$\alpha$ network (GHN) at BBSO and Kanzelhoehe Solar Observatory.


\bibliographystyle{aa}
\bibliography{references}
\end{document}